# Symmetry breaking of Pancharatnam–Berry phase using non-axisymmetric meta-atoms

**Baifu Zhang,[1][*] Yan Wang,[1] Zhixing Huang,[1] Huafeng Li,[1] Ji Xu,[2] and Jianping Ding[3],[4]**

[1]*School of Electronic and Optical Engineering, Nanjing University of Science and Technology, Nanjing 210094, China*
[2]*College of Electronic and Optical Engineering & College of Flexible Electronics (Future Technology), Nanjing University of Posts and Telecommunications, Nanjing 210023, China*
[3]*Collaborative Innovation Center of Advanced Microstructures and School of Physics, Nanjing University, Nanjing 210093, China*
[4]*jpding@nju.edu.cn*
[*]*zhangbf@njust.edu.cn*

**Abstract:** The Pancharatnam–Berry (PB) phase in metasurfaces obeys the symmetry restriction, according to which the PB phases of two orthogonal circularly polarized waves are the same but with opposite signs. Here, we reveal a general mechanism to break the axisymmetry of meta-atoms in order to break the PB-phase symmetry restriction. As a proof of concept, we designed a novel meta-atom with a QR-code structure and successfully demonstrated circular-polarization multiplexing metasurface holography. This study provides a fundamentally new understanding of the PB phase and opens a path for arbitrary wavefront engineering using asymmetric electromagnetic structures.

## 1. Introduction

Metasurfaces, as ultrathin electromagnetic (EM) functional layers, have emerged as a comprehensive and compact platform for wavefront engineering in the last decade [1,2]. The overall wavefront engineering by a metasurface results from the linear superposition of EM waves modulated by each meta-atom, which acts as the unit cell of the metasurface. The conventional phase-modulation mechanisms of a meta-atom include the dynamic phase and Pancharatnam–Berry (PB) phase (or geometric phase). The dynamic phase originates from the accumulated phases of an EM wave propagating in a meta-atom; the phase accumulation depends on the refractive index, geometry, and other meta-atom parameters [3,4]. The PB phase arises from the spin–orbit coupling of photons in a meta-atom, as a function of the rotation angle of the anisotropic EM structures (including elliptical structures and rectangular structures) [5–7]. For example, when a circularly polarized EM wave propagates through such structures with a rotation angle $\theta$, the transmitted cross-polarized component adopts the so-called PB phase ($\Phi$). Conventionally, $\Phi = \pm 2\theta$, where the sign ± is determined by the chirality of the EM wave; for instance, the left- and right-circular polarizations (LCP and RCP, respectively). Unlike the dynamic phase, the PB phase is symmetric for both the orthogonal circular polarizations. In other words, the PB phases of the LCP and RCP have the same absolute value but opposite signs. Recently, Xie et al. studied the rotational symmetry of meta-atoms and demonstrated high-order PB phases, which are equivalent to several times the rotation angle (rather than just two times) [8]. However, the PB phase of these meta-atoms remains symmetric. Therefore, achieving polarization-decoupled EM wavefront engineering using only the geometric phase is difficult.

In recent years, some researchers attempted to break the symmetry of the PB phase. For example, by introducing nonlinear effects, the PB phase of transmitted harmonic waves can be rewritten in the form of $\pm(n\pm1)\theta$, where n is the order of harmonic generation [9–11]. However, this method cannot be used to engineer the PB phases of orthogonal circular polarizations arbitrarily. In the case of linear processes, Bai et al. proposed to combine the Aharonov–Anandan and PB phases to break the symmetry limitation of the PB phase [12]. This proposed method can be applied in EM wave engineering metasurfaces [13,14]. Chen et al. proposed a type of planar chiral meta-atom to perform local phase manipulations in metasurfaces and realized a controlled spin decoupling of the orthogonal circularly polarized waves. [15]. The studies reported to date trace the origins of asymmetric PB phases from different physical mechanisms and demonstrate polarization-decoupled PB phase modulation. However, the identification of a general fundamental factor that can break the symmetry limitation of the PB phase for orthogonal circularly polarized waves is crucial for understanding the relationship between meta-atom structures and their corresponding EM wavefront engineering properties.

In this study, we analytically demonstrated that the symmetry restriction of the PB phase originates from the axisymmetry of EM structures for the first time to the best of our knowledge. In other words, from the geometrical and topological point of view, a non-axisymmetric meta-atom topology naturally leads to symmetry breaking of the PB phases of two orthogonal circularly polarized waves, although different geometric structures may have different polarization-decoupling mechanisms. Further, we analytically established the relationship between the Jones matrices based on the circular- and linear-polarization base vectors. Moreover, we designed and investigated non-axisymmetric meta-atoms with QR-code structures to verify the breaking of the symmetry restriction of the PB phase. As a proof of concept, we designed a circularly polarized multiplexed hologram using a metasurface consisting of QR-code meta-atoms and numerically demonstrated our proposed theory. The results of this study provide a fundamentally new understanding of the PB phase and light–matter interactions in nanophotonics and can thus promote more advanced metasurface- and metamaterial-based applications.

## 2. Theory of PB phase for non-axisymmetric meta-atoms

First, we derive a general form of the PB phase for a meta-atom with an arbitrary topology, e.g., a non-axisymmetric structure. Note that the structure of a three-dimensional (3D) meta-atom is also called a "non-mirror symmetric" structure instead of a non-axisymmetric structure. However, considering that the topologies of meta-atoms usually vary in the top-view plane and remain unchanged in the height dimension, we perform the analysis using the "non-axisymmetry" concept for simplicity but without loss of generality. For an EM wave incident on a meta-atom, the relationship between the transmitted and incident waves can be established using the Jones matrix. Usually, we use linearly polarized base vectors to describe the incident and transmitted EM waves as follows:

$$E_{in} = a_1 \hat{x} + a_2 \hat{y} = \begin{pmatrix} a_1 \\ a_2 \end{pmatrix}, \tag{1}$$

$$E_{out} = J E_{in} = \begin{pmatrix} J_{11} & J_{12} \\ J_{21} & J_{22} \end{pmatrix} \begin{pmatrix} a_1 \\ a_2 \end{pmatrix}, \tag{2}$$

where $E_{in}$ and $E_{out}$ are the incident and transmitted EM waves, respectively; $\hat{x}$ and $\hat{y}$ are the base vectors of the $x$ and $y$ linear polarizations, respectively; $a_1$ and $a_2$ are the corresponding complex amplitudes with $a_1^2 + a_2^2 = 1$ as the normalization condition; and $J$ is the Jones matrix for linear-polarization base vectors. For an arbitrary meta-atom structure, the four elements of $J$ are all non-zero values. However, for an axisymmetric structure, $J_{21} = J_{12} = 0$ (details in the Appendix), which suggests that a linear-polarization-decoupled metasurface can be designed using axisymmetric meta-atoms as discussed in our previous work [16].

Introducing circular-polarization base vectors to the above equations, we obtain

$$E_{in} = b_1 \hat{L} + b_2 \hat{R} = \begin{pmatrix} b_1 \\ b_2 \end{pmatrix}, \tag{3}$$

$$E_{out} = SE_{in} = \begin{pmatrix} S_{11} & S_{12} \\ S_{21} & S_{22} \end{pmatrix} \begin{pmatrix} b_1 \\ b_2 \end{pmatrix}, \tag{4}$$

where $\hat{L}$ and $\hat{R}$ are the base vectors of LCP and RCP, respectively; $b_1$ and $b_2$ are the corresponding complex amplitudes with $b_1^2 + b_2^2 = 1$ as the normalization condition; and $S$ is the Jones matrix for circular-polarization base vectors. If we further consider the relationship between two sets of orthogonal base vectors as $\hat{L} = \frac{\sqrt{2}}{2}(\hat{x} - i\hat{y})$ and $\hat{R} = \frac{\sqrt{2}}{2}(\hat{x} + i\hat{y})$, and conduct some analytical derivations (details in the Appendix), then we can rewrite the $S$ matrix in the form of $J$ matrix as follows:

$$\begin{cases} S_{11} = \frac{1}{2}[(J_{11} + J_{22}) - i(J_{12} - J_{21})] \\ S_{21} = \frac{1}{2}[(J_{11} - J_{22}) - i(J_{12} + J_{21})] \\ S_{12} = \frac{1}{2}[(J_{11} - J_{22}) + i(J_{12} + J_{21})] \\ S_{22} = \frac{1}{2}[(J_{11} + J_{22}) + i(J_{12} - J_{21})] \end{cases}. \tag{5}$$

Further, if the meta-atom rotates by an angle $\theta$, then the Jones matrix is expressed as

$$J(\theta) = R(-\theta)JR(\theta) = \begin{pmatrix} \cos\theta & -\sin\theta \\ \sin\theta & \cos\theta \end{pmatrix} \begin{pmatrix} J_{11} & J_{12} \\ J_{21} & J_{22} \end{pmatrix} \begin{pmatrix} \cos\theta & \sin\theta \\ -\sin\theta & \cos\theta \end{pmatrix}. \tag{6}$$

We substitute Eqs. (2) and (6) into Eq. (5), and finally obtain the most general form of the PB phase for an arbitrary meta-atom using circular-polarization base vectors as follows:

$$\begin{cases} S_{11}(\theta) = \frac{1}{2}[(J_{11} + J_{22}) - i(J_{12} - J_{21})] \\ S_{21}(\theta) = \frac{1}{2}[(J_{11} - J_{22}) - i(J_{12} + J_{21})]e^{-i2\theta} \\ S_{12}(\theta) = \frac{1}{2}[(J_{11} - J_{22}) + i(J_{12} + J_{21})]e^{i2\theta} \\ S_{22}(\theta) = \frac{1}{2}[(J_{11} + J_{22}) + i(J_{12} - J_{21})] \end{cases}. \tag{7}$$

Compared to the linear Jones matrix $J(\theta)$, $S(\theta)$ intuitively exhibits numerous important properties. First, the diagonal elements $S_{11}$ and $S_{22}$ remain unchanged even if the meta-atom is rotated. Second, the anti-diagonal elements $S_{21}$ and $S_{12}$ for the transmitted cross-polarizations are naturally symmetry-broken. In other words, the cross-polarization transmissions of the incident circularly polarized EM waves are inherently asymmetric, which is contrary to the traditional view. PB phases are subjected to symmetry restrictions in conventional meta-atoms such as those with rectangular, elliptical, and other structures, which have been widely studied in previous works, because for the aforementioned axisymmetric topologies, $J_{12} = J_{21} = 0$. Thus, the asymmetric elements $S_{21}(\theta)$ and $S_{12}(\theta)$ degenerate into the most known forms with a symmetric PB phase:

$$\begin{cases} S_{21}(\theta) = \frac{1}{2}(J_{11} - J_{22})e^{-i2\theta} \\ S_{12}(\theta) = \frac{1}{2}(J_{11} - J_{22})e^{i2\theta} \end{cases}. \tag{8}$$

The conventional symmetry restriction of the PB phase reportedly originates from the axisymmetry of meta-atoms, and thus breaking the axisymmetry of the meta-atoms intrinsically leads to breaking of the PB-phase symmetry. Notably, instead of the structural symmetry, the optical mode (electromagnetic field) inside the meta-atom structure should be adopted as a more general physical quantity in this case; however, these two factors are usually consistent with each other. Thus, for simplicity, we perform our analysis from the structural point of view.

Next, we rewrite $S_{21}(\theta)$ and $S_{12}(\theta)$ in Eq. (7) in a more intuitive form. If $S_{21}$ and $S_{12}$ simplify to $A_{21}e^{i\varphi_{21}}$ and $A_{12}e^{i\varphi_{12}}$, respectively, then $S_{21}(\theta)$ and $S_{12}(\theta)$ can be expressed as:

$$\begin{cases} S_{21}(\theta) = (A_{21}e^{i\varphi_{12}})e^{-i(2\theta-\Delta\varphi)} \\ S_{12}(\theta) = (A_{12}e^{i\varphi_{12}})e^{i2\theta} \end{cases}, \quad (9)$$

where $A_{12}$ and $A_{21}$ are the amplitudes of $S_{21}$ and $S_{12}$, respectively; $\varphi_{21}$ and $\varphi_{12}$ are the argument angles of $S_{21}$ and $S_{12}$, respectively; and $\Delta\varphi = \varphi_{21} - \varphi_{12}$. The symmetry of PB phase is broken by the key phase $\Delta\varphi$.

## 3. Schematic of the non-axisymmetric meta-atom with QR-code structure

In order to break the symmetry restriction of the PB phase, the meta-atom should break the axisymmetry in the topology. Here, we design a novel dielectric meta-atom with a QR-code structure to eliminate any geometric symmetry as shown in Fig. 1. This type of QR-code topology, which was investigated in our previous study on perfect metamaterial absorbers, possesses a large number of degrees of freedom and can thus provide rich properties for wavefront engineering [17]. In this study, the meta-atom, with a period $P$ of 800 nm, consists of a silica substrate and arrays of square silicon pillars that appear as a QR code. Each QR code consists of 5 × 5 pixels, and each pixel is randomly set as either a square silicon pillar (length: 100 nm; height: 1400 nm) or air. Because the generated QR-code topology is completely random, it easily and thoroughly breaks the axisymmetry.

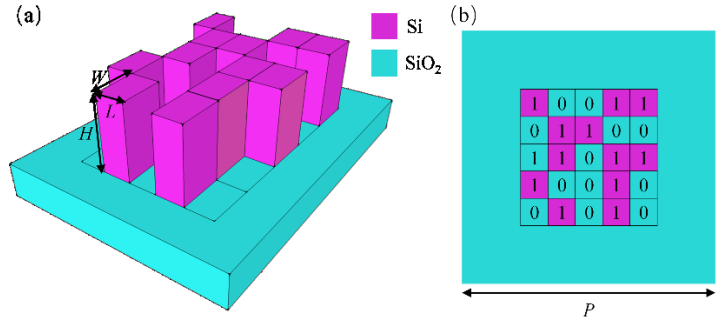

Fig. 1 Schematic of the non-axisymmetric meta-atom with QR-code structure: (a) 3D view and (b) top view. The parameters are set as $P$ = 800 nm, $L$ = $W$ = 100 nm, and $H$ = 1400 nm.

We conducted full-wave 3D finite-difference time-domain (FDTD) simulations to design and investigate the QR-code meta-atom structures. In the simulations, we set the refractive index of the silica substrate as 1.45, and derived the dielectric constant values of silicon by fitting Palik's experiment data to the Drude–Lorentzian model [18,19]. For the meta-atom simulation, we used the period boundaries in the horizontal direction and perfectly matched layers (PMLs) in the longitudinal direction. By randomly setting each pixel of the QR-code structure, we obtained a variety of non-axisymmetric meta-atoms and numerically calculated their parameters. All the simulated meta-atoms show asymmetric PB phases, which are discussed in detail in the next section.

## 4. Breaking the symmetry restriction of PB phase using non-axisymmetric meta-atoms

We designed and numerically evaluated 13,000 meta-atom structures with randomly arrayed 5×5 QR-code nanopillars. All the QR-code meta-atoms show decoupled PB phases for LCP and RCP. We selected structures with transmission coefficients higher than 0.7 to ensure efficient polarization-multiplexing applications, and the PB phases of these selected structures are plotted in Fig. 2(a). Evidently, the PB phases of the LCP and RCP cover the 0–2π range and break the symmetry restriction, consistent with the predictions of Eqs. (7) and (9). For simplicity but without loss of generality, we do not rotate the meta-atoms in Fig. 2, indicating that the rotation angle $\theta$ is zero. The corresponding results suggest that the polarization-decoupled PB phases, which originate from non-axisymmetric structures, are new degrees of freedom for engineering EM wavefronts. In addition, we can independently engineer the PB phases for orthogonal circular polarizations without rotating the meta-atoms in the full range from 0 to $2\pi$.

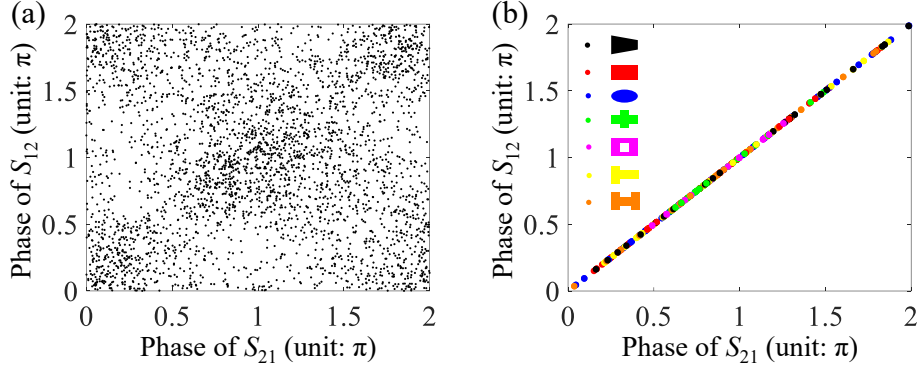

Fig. 2 PB phases of different meta-atom structures. (a) PB phases of the QR-code structures shown in Fig. 1. (b) PB phases of different axisymmetric structures.

We investigated a variety of axisymmetric meta-atom structures as a control group, and their PB phases are plotted in Fig. 2(b). All axisymmetric meta-atoms have the same period (800 nm) and height (1400 nm) as the non-axisymmetric ones, but different topologies in the top view. All PB phases are located on the $y = x$ line, irrespective of the meta-atom topology (such as rectangle, ellipse, trapezoid, cross, T-shape, H-shape, or hollow rectangle), implying that all PB phases of the two orthogonal circular polarizations have the same value. This is the symmetry restriction of conventional PB phases, which is predicted in Eq. (8), and originates from axisymmetric structures.

From Fig. 2, we can verify our analytical predictions of the PB phases; that is, decoupling of the PB phases can be achieved under a broken axial symmetry. Furthermore, the decoupled PB phase, independent of the rotation angle of the meta-atom, provides a new degree of freedom in wavefront engineering. As a proof of concept, we designed a circular-polarization multiplexing metasurface hologram based on our proposed QR-code meta-atoms.

The design of the circular-polarization-decoupled metasurface hologram is illustrated in a flowchart in Fig. 3. First, we calculated the phase-only computer-generated holograms (CGHs) for left- and right-circular polarized incident light beams using the Gerchberg–Saxton (GS) algorithm [20]. We coded two different images for the two orthogonal polarizations and obtained two output holographic phase diagrams after the GS iterations. Second, we performed an eight-level phase quantization to rewrite the two phase diagrams into two phase matrices. In this process, we selected $8 \times 8$ meta-atom structures from Fig. 2(a) to establish the eight-level phase library, which was then searched to identify the meta-atom structures that fulfilled both the phase requirements for the LCP and RCP holograms; to identify these structures, a one-by-one element search of the phase matrices was performed. Finally, we simulated the desired metasurface hologram with 64 types of QR-code structures. Due to the limited computing resources, we could simulate a metasurface with $101 \times 101$ periods as a proof of concept, which successfully demonstrated our theory. As the period was set to 800 nm, the metasurface hologram could cover an area of $80.8 \times 80.8 \ \mu m^2$.

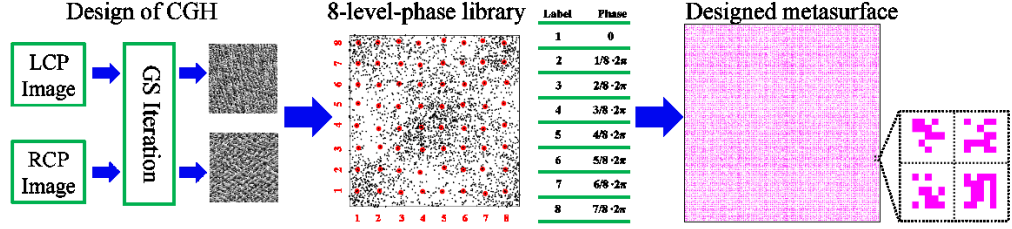

Fig. 3 Design flowchart of the circular-polarization-decoupled metasurface hologram.

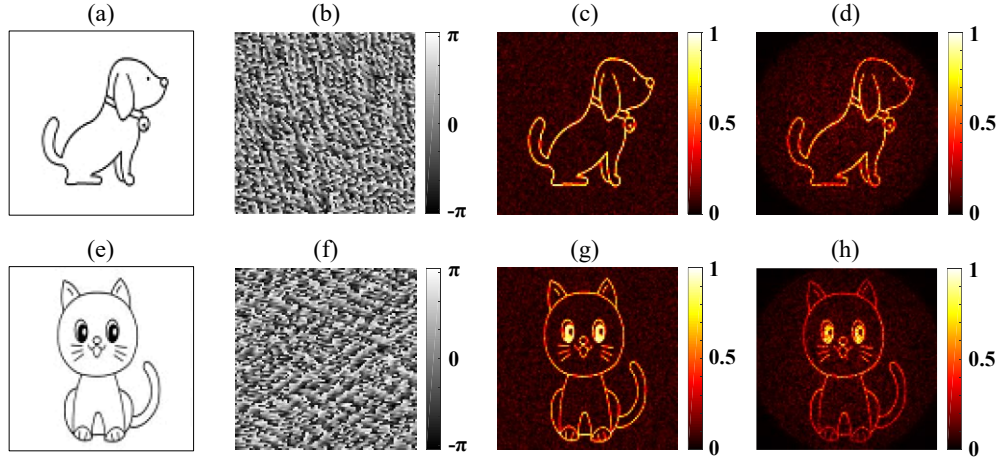

Fig. 4. Circular-polarization-decoupled metasurface holography. (a) and (e) are the original images for LCP and RCP illuminations, respectively. (b) and (f) correspond to the phase distributions of the calculated holograms. (c) and (g) are the images of the theoretical reconstructions. (d) and (h) correspond to the reconstructed images of the simulated metasurface hologram.

The designed metasurface hologram was simulated using 3D FDTD and is shown in Fig. 3, and the reconstructed images for LCP and RCP are presented in Fig. 4. Figures 4 (a) and (e) show the 101 × 101-pixel dog and cat images as the original images, which are illuminated by left- and right-circularly polarized light, respectively. Thus, the images reconstruct using the PB phases incorporate the cross-polarization components RCP and LCP. The phase-only CGHs of the two orthogonal polarizations, simulated using the GS algorithm, are show in Figs. 4(b) and (f). As the control group, the images of the theoretical reconstructions are shown in Figs. 4(c) and (g). The reconstructed images of the simulated metasurface hologram, which was illuminated by different circularly polarized EM waves, are plotted in Figs. 4(d) and (h). We can see that the reconstructed images of the metasurface hologram show good agreement with the original and theoretically reconstructed images, thereby confirming the feasibility of decoupling circularly polarized light via wavefront engineering. This result demonstrates that non-symmetric meta-atom structures can indeed break the symmetry restriction of PB phases and can provide new insights into metasurfaces and metamaterials to promote their application in various fields.

## 5. Conclusion

In conclusion, we theoretically established a general formula to describe the wavefront engineering property of a meta-atom using the Jones matrix and circular-polarization base vectors. The analytical result indicates that the non-axisymmetry of meta-atom leads to the breaking of the symmetry restriction of the PB phase. As an illustrative example, we designed a novel QR-code meta-atom to break the topological axisymmetry and realize polarization-decoupled PB-phase engineering. Further, we designed a circularly polarized multiplexed metasurface hologram using our proposed QR-code meta-atoms to reveal the potential of symmetry-broken PB phases in various applications in metasurfaces and metamaterials. We believe that this study will expand the understanding of the PB phase in a fundamental way and will facilitate the development of design methodologies for EM structures, which can be used for arbitrary wavefront engineering.

**Appendix A: Derivation of Jones matrix *S***

When a left-circularly polarized light beam is incident on an arbitrary meta-atom, the transmitted (or reflected) light is

$$E_{out} = SE_{in} = \begin{pmatrix} S_{11} & S_{12} \\ S_{21} & S_{22} \end{pmatrix}\begin{pmatrix} 1 \\ 0 \end{pmatrix} = S_{11}\hat{L} + S_{21}\hat{R} = S_{11}\frac{\sqrt{2}}{2}(\hat{x} - i\hat{y}) + S_{21}\frac{\sqrt{2}}{2}(\hat{x} + i\hat{y}) \quad (A1)$$

Similarly, when a right-circularly polarized light beam incidents, the output light field is expressed as:

$$E_{out} = SE_{in} = \begin{pmatrix} S_{11} & S_{12} \\ S_{21} & S_{22} \end{pmatrix}\begin{pmatrix} 0 \\ 1 \end{pmatrix} = S_{12}\hat{L} + S_{22}\hat{R} = S_{12}\frac{\sqrt{2}}{2}(\hat{x} - i\hat{y}) + S_{22}\frac{\sqrt{2}}{2}(\hat{x} + i\hat{y}) \quad (A2)$$

We can rewrite the above equations using linear-polarization base vectors. For a left-circularly polarized incident light beam, the transmitted electric field is given by

$$E_{out} = JE_{in} = \begin{pmatrix} J_{11} & J_{12} \\ J_{21} & J_{22} \end{pmatrix}\begin{pmatrix} \frac{\sqrt{2}}{2} \\ -\frac{\sqrt{2}}{2}i \end{pmatrix} = \frac{\sqrt{2}}{2}(J_{11} - iJ_{12})\hat{x} + \frac{\sqrt{2}}{2}(J_{21} - iJ_{22})\hat{y} \quad (A3)$$

Similarly, for a right-circularly polarized incident light beam, we obtain

$$E_{out} = JE_{in} = \begin{pmatrix} J_{11} & J_{12} \\ J_{21} & J_{22} \end{pmatrix}\begin{pmatrix} \frac{\sqrt{2}}{2} \\ \frac{\sqrt{2}}{2}i \end{pmatrix} = \frac{\sqrt{2}}{2}(J_{11} + iJ_{12})\hat{x} + \frac{\sqrt{2}}{2}(J_{21} + iJ_{22})\hat{y} \quad (A4)$$

The physical processes represented by the aforementioned equations remain identical irrespective of using a circular- or linear-polarization base vector. Thus, we solve Eqs. (A1)–A(4) simultaneously and establish the analytical relationship between *J* and *S* as follows:

$$\begin{cases} S_{11} = \frac{1}{2}[(J_{11} + J_{22}) - i(J_{12} - J_{21})] \\ S_{21} = \frac{1}{2}[(J_{11} - J_{22}) - i(J_{12} + J_{21})] \\ S_{12} = \frac{1}{2}[(J_{11} - J_{22}) + i(J_{12} + J_{21})] \\ S_{22} = \frac{1}{2}[(J_{11} + J_{22}) + i(J_{12} - J_{21})] \end{cases} \quad (A5)$$

**Appendix B: Axisymmetry of meta-atoms**

For an *x*-polarized incident light beam, the output light wave is given by

$$\begin{pmatrix} J_{11} & J_{12} \\ J_{21} & J_{22} \end{pmatrix}\begin{pmatrix} 1 \\ 0 \end{pmatrix} = J_{11}\hat{x} + J_{21}\hat{y} \quad (B1)$$

If a mirror operation is performed on the meta-atom along the symmetric axis (for example, the *x*-axis), then the Jones matrix $J'$ becomes $\begin{pmatrix} J'_{11} & J'_{12} \\ J'_{21} & J'_{22} \end{pmatrix}$, and the output becomes

$$\begin{pmatrix} J'_{11} & J'_{12} \\ J'_{21} & J'_{22} \end{pmatrix} \begin{pmatrix} 1 \\ 0 \end{pmatrix} = J'_{11}\hat{x} + J'_{21}\hat{y} \tag{B2}$$

According to the property of mirror operation, the *y*-polarization component effectively undergoes a phase retardation of $e^{i\pi}$, and thus, we obtain $J_{11}\hat{x} = J'_{11}\hat{x}$, $J_{21}\hat{y} = J'_{21}\hat{y} \cdot e^{i\pi}$, which implies $J'_{11} = J_{11}$, $J'_{21} = -J_{21}$.

Similarly, in the case of the *y*-polarization incident light beam, we can obtain $J'_{12} = -J_{12}$, $J'_{22} = J_{22}$. Thus, the relationship between $J$ and $J'$ can be expressed as

$$J' = \begin{pmatrix} J'_{11} & J'_{12} \\ J'_{21} & J'_{22} \end{pmatrix} = \begin{pmatrix} J_{11} & -J_{12} \\ -J_{21} & J_{22} \end{pmatrix} \tag{B3}$$

For an axisymmetric meta-atom, mirror operation on its symmetric axis does not change its geometry. Thus, we obtain following equation:

$$J' = \begin{pmatrix} J'_{11} & J'_{12} \\ J'_{21} & J'_{22} \end{pmatrix} = \begin{pmatrix} J_{11} & -J_{12} \\ -J_{21} & J_{22} \end{pmatrix} = \begin{pmatrix} J_{11} & J_{12} \\ J_{21} & J_{22} \end{pmatrix} = J \tag{B4}$$

which indicates $J_{12} = -J_{12} = 0$, $J_{21} = -J_{21} = 0$. Thus, the Jones matrix $J$ of the axisymmetric meta-atoms is simplified to

$$J_{axis-symmetry} = \begin{pmatrix} J_{11} & 0 \\ 0 & J_{22} \end{pmatrix} \tag{B5}$$

Further, if the axisymmetric meta-atom rotates by an angle $\theta$, then the corresponding Jones matrix $S(\theta)$ in Eq. (7) is simplified to

$$S_{axis-symmetry}(\theta) = \frac{1}{2}\begin{pmatrix} J_{11} + J_{22} & (J_{11} - J_{22})e^{i2\theta} \\ (J_{11} - J_{22})e^{-i2\theta} & J_{11} + J_{22} \end{pmatrix} \tag{B6}$$

**Funding.** National Key Research and Development Program of China (2022YFA1404800, 2018YFA0306200); National Natural Science Foundation of China (11922406, 91750202);

**Acknowledgments.** We thank Mr. Yong Chen for the fruitful discussions and his help provided during this work.

**Disclosures.** The authors declare no conflicts of interest.

**Data availability.** Data underlying the results presented in this paper are not publicly available at this time but may be obtained from the authors upon reasonable request.